# Kinesin Is An Evolutionarily Fine-Tuned Molecular Ratchet-And-Pawl Device of Decisively Locked Direction

Zhisong Wang[1*], Min Feng[2], Wenwei Zheng[1], Dagong Fan[1]

[1]Institute of Modern Physics and Applied Ion Beam Physics Laboratory, Fudan University, Shanghai 200433, China

[2]Genome Damage and Stability Centre, University of Sussex, East Sussex BN1 9RQ, U.K.

*Corresponding author (E-mail: wangzs@fudan.edu.cn)

**ABSTRACT**

Conventional kinesin is a dimeric motor protein that transports membraneous organelles towards plus end of microtubules (MTs). Individual kinesin dimers show steadfast directionality and hundreds of consecutive steps, yet the detailed physical mechanism remains unclear. Here we compute free energies for the entire dimer-MT system for all possible interacting configurations by taking full account of molecular details. Employing merely first principles and several measured binding and barrier energies, the system-level analysis reveals insurmountable energy-gaps between configurations, asymmetric ground state caused by mechanically lifted configurational degeneracy, and forbidden transitions ensuring coordination between both motor domains for alternating catalysis. This wealth of physical effects converts a kinesin dimer into a molecular ratchet-and-pawl device, which determinedly locks the dimer's movement into MT plus end and ensures consecutive steps in hand-over-hand gait. Under a certain range of extreme loads however, the ratchet-and-pawl device becomes defect but not entirely abolished to allow consecutive back-steps. This study yielded quantitative evidence that kinesin's multiple molecular properties have been evolutionarily adapted to fine-tune the ratchet-and-pawl device so as to ensure the motor's distinguished performance.

**INTRODUCTION**

Conventional kinesin(1) has two dimerized motor domains (also called "head" regions) which each is made of a catalytic core domain containing the nucleotide- and microtubule-binding sites and a neck domain sufficient for dimerizing the two motor domains. Conventional kinesin is distinct from other microtubule-based motor proteins in that individual kinesin dimers are steadfast in their self-chosen direction(2-4), and can run hundreds of consecutive steps before falling off the microtubule (MT)(5). These unusual motor capabilities are unique properties of dimerized motor domains. A monomeric protein of kinesin superfamily, KIF1A was found to diffuse back and forth along MT under a bias towards the plus end(6). However the monomer is incapable of processive movement, and its average direction is easily reversed by an opposing loads of less than a pN(6). In contrast, a dimer of conventional kinesin can evidently retain its directionality not in an average sense but in a determined manner(2-4). Depending on ATP (Adenosine Triphosphate) concentrations, an opposing force of 5 – 8 pN(3) brings conventional kinesin to a complete halt. Then the dimer's center-of-mass undergoes balanced back-and-forth single-step movement but develops no consecutive back-steps till the whole dimer falls off MT(2-4). As for kinesin's processvity (i.e. long run of consecutive steps), kinetic studies(7-9) suggested that the two heads of a kinesin dimer alternately hydrolyze ATP to ensure the dimer's continual runs in an acrobatic head-over-head gait(10). The kinetic model of alternating head catalysis requires a molecular mechanism for head-head coordination(7-9), which however remains mysterious thus far.

The first ~14 residues of the neck bridging the catalytic core and the subsequent coiled-coil dimerization domain is termed neck linker(1), which has been identified as a key determinant of kinesin's directionality(1,11-13). The neck linker is immobilized on to the catalytic core and extends towards the MT plus end when the catalytic core is both MT- and ATP-bound(11). This conformational change is termed "neck linker zippering"(11). A zippered neck linker is reverted to a random conformation upon γ-phosphate release(11). In their mechanically controlled access model(1), Vale and Milligan indicated that the neck linkers of a dimer must be overstretched for both heads to bind MT simultaneously, and neck linker zippering at a standing head enables the other diffusing head to reach the nearby binding site to the MT plus end but hinders access to the posterior site. It was later found that the identified conformational change has a free-energy gain of merely ~ 1.2 $k_B T$ ($k_B$ is Boltzmann constant, and $T$ absolute temperature)(13), which appears to be insufficient to account for the kind of robust directionality shown by kinesin dimers(14). A new twist is added to the issue of kinesin's directionality by a recent study(4), in which quickly applying super-stall loads of 10-15 pN caused a kinesin dimer to walk consecutive backward steps. Understanding of these seemingly contradictory results presents a challenge to the field of kinesin study.

Because conventional kinesin's robust directionality and high processivity are both properties beyond individual monomers, the underlying mechanisms likely involve the entire system of dimerized motor domains together with MT. To close in on the synergetic mechanisms we have computed free energies for all possible interacting configurations for the whole dimer-MT system by taking full account of their atomic details. The computed configurational energies form a unique hierarchy, which naturally imposes non trivial constraints on transitions between the configurations in a way similar to hydrogen molecule's electronic levels in defining the absorption and emission spectra. Arising from the system-level energetic hierarchy, the transition rules lead straightforwardly to a wealth of molecular mechanisms for coordination between both motor domains in their mechanochemical activities. Such a system-level analysis thereby reveals a unified mechanism for sturdy unidirectionality and high processivity of conventional kinesin. This method of system-level energetic analysis is different from theoretical methods previously applied to kinesin(5,15-21).

**THEOTY AND METHOD**
**Computation of energies for kinesin-MT binding configurations**

The total free energy for a dimer-MT configuration is a sum of free energy of the linker chains ($F$), free-energy gain by linker zippering ($U_Z$), and head-MT binding energies ($U_B$). Experiments(13,22) yielded $U_Z$ = -1.2 $k_B T$, $U_B$ = -15 $k_B T$ for ADP-bound heads, and $U_B$ = -19 $k_B T$ for ATP-bound or nucleotide-free heads. The Helmholtz free energy of the linker chains in a double-headed binding configuration can be calculated using a formula derived from worm-like chain model for polymers

$$F(l_{eff}, d_{eff}) = (k_B T)\left(\frac{l_{eff}}{l_p}\right)\left[\frac{(d_{eff}/l_{eff})^2(3 - 2d_{eff}/l_{eff})}{4(1 - d_{eff}/l_{eff})}\right] . \tag{1}$$

Here $l_p$ is persistence length of the linker peptides. $d_{eff}$ is the extension of both linkers required for forming the dimer-MT double-bindings, $l_{eff}$ is the portion of linker contour length available for spanning $d_{eff}$.

In a dimmer-MT double-binding configuration both motor heads bind MT in similar orientation along a single protofilament (23,24). A geometry analysis yields

$$d_{eff} = d - (m_Z - 2m_T)a , \tag{2}$$

$$l_{eff} = 2l_N - m_T l_T - m_Z l_Z . \tag{3}$$

Here $d$ is protofilament lattice spacing of MT (i.e. binding site period), $l_N$ is contour length of a linker chain. $l_Z$ is the portion of contour length zippered to the catalytic core, and $a$ is projection of $l_Z$ in parallel to MT. We take $d$ = 8.2 nm and $a = 0.95 l_Z$ following structural studies(25). A subtlety is that zippering at the leading head will force the linker chain to form a half loop, which takes a small but finite portion of contour length $l_T$ (26). In eq. 2 and 3 $m_T$ is number of half-loops, and $m_Z$ is number of zippered linkers. Persistence length and looping property of the linker peptides were determined from an atomic computation (See APPENDIX), which yields $l_T$ = 0.36 nm and $l_p$ = 0.8 nm. The zippered length $l_Z$ will be deduced from measured barriers for kinesin's steps in the section of RESULTS AND DISCUSSIONS.

We derived eq. 1 from an interpolation formula suggested in Ref. (27,28) for force-extension relationship of a worm-like chain:

$$f = (k_B T / l_p)\left[(1 - z/l)^{-2}/4 - 1/4 + z/l\right]. \quad (4)$$

In this equation $f$ is the pulling force applied at the ends of the polymer chain, $z$ is the average end-to-end extension, and $l$ is the polymer's contour length. Substituting $z$ by $d_{eff}$ and $l$ by $l_{eff}$, and integrating the interpolation equation yielded Gibbs free energy for the linker peptides, which in turn yielded the Helmholtz free energy (eq. 1).

Our choice of the worm-like chain model, and specifically the interpolation formula to describe stretched states of the neck linkers in double-headed kinesin-MT binding configurations is based on the following reasons. In a double-headed binding configuration the two linker peptides of total ~28 amino acid residues span the protofilament lattice spacing of ~ 8 nm, which is an overstretching situation of $z/l$ > 0.75. In this overstretching regime different polymer models predict drastically differing force-extension curves for a polymer chain (see Fig. 2.15 of ref.(29)). As $z/l \to 1$, the worm-like chain model predicts $f \propto (1 - z/l)^{-1}$ whilst the freely jointed chain model predicts $f \propto (1 - z/l)^{-2}$. Accuracy of the worm-like chain model in quantitative description of stretched DNA and polypeptides has been proved by single-molecule experiments(27,28,30-32). Previous studies(27,28) found however that predictions of the freely jointed chain model deviate from measured force-extension curves in the overstretching regime of $z/l$ > 0.75. The Gaussian chain model, broadly used because of its relatively simple mathematical form, is generally not applicable to overstretched polymers (see Fig. 2.15 of ref. (29)). Previous studies(33) also found that applicability of the Gaussian chain model to short peptides is questionable.

The interpolation formula, being an approximate equation though, has been found to reproduce very well measured force-extension curves of single DNA or protein molecules(27,28,30,31). The agreement is satisfactory also for mechanical stretching and breaking of single protein domains which each has a contour length of merely ~ 30 nm (see figure 3 of Ref. (30)). The proved applicability to short peptides justifies our choice of the interpolation formula as a basis for studying stretched states of kinesin's neck linkers. Previous studies(28) showed that the interpolation equation deviates from the exact solution of the worm-like chain model by ~ 10% at $z/l$ = 0.5, and the error systematically decreases as $z/l$ increases into the overstretching regime of $z/l >$ 0.75 relevant to the present study.

Following structural studies(25) we use $n_N$ = 14 amino acid residues for kinesin's linker length (i.e. $l_N = n_N \times 0.36$ nm). We however note that the coiled coils can unwind to increase effective length of the linker. Recent studies(34) have ruled out complete melting of the first heptad repeat, which would otherwise add 7 amino acid residues to effective linker length and render defect the maximum directional preference. Nevertheless, partial unwinding at the beginning 1-2 residues of the coiled coil is likely in double-headed kinesin-MT binding configurations, in which a mechanical strain well above 20 pN is developed along the linker chains as estimated using eq. 1. Such an amount of force is sufficient to break individual molecular contacts within heptad repeats(30,31). For double-binding dimer-MT configurations the effective linker length therefore might range between 14 - 16 residues (marked in Fig. 1 B).

For the cases in which a constant force is applied to the stalk domain adjacent to the neck coiled coils, extensions of both linkers are determined by balancing forces at the coiled coil domain. Internal Helmholtz free energies of individual neck linkers are then combined with the contribution due to the external force to yield Gibbs free energy(33) for the neck linkers. Under an opposing load the neck linker adjacent to the front head in a double-headed binding configuration is more extended than the linker adjacent to the rear head. Consequently the forces inflicted upon the two MT-bound heads by their adjacent neck linkers are different, and are given by derivatives of the Helmholtz free energies of the respective linker peptides. The forces felt by individual heads will be used in considering load-dependence of enzymatic rates of motor domains (see the following subsection). In single-headed binding configurations the standing head alone bears the effect of the external load.

**Kinetic Monte Carlo simulation of kinesin's walking dynamics**

Because of stochastic nature of both enzymatic processes and head diffusion, the kinetic Monte Carlo method(35) is suitable to simulation of Kinesin's walking behavior. The kinetic Monte Carlo simulation follows time evolution of the kinesin-MT system as it undergoes transitions between different configurations as driven by cycles of ATP hydrolysis at the two heads. Within the framework of the kinetic Monte Carlo method, diffusion of heads is not treated explicitly. Rather a rate for a diffusing head to bind MT was calculated by considering the geometrical and energetic differences between the initial single-headed binding configuration and the final configuration of double-headed bindings. This rate for random search-and-binding then was used in the kinetic Monte Carlo algorithm(35-37) to calculate transitions from single-headed to double-headed binding configurations of the kinesin-MT system. More specifically, head-MT binding is assumed to occur once the diffusive head encounters a binding site of MT, and the search-and-binding rate can then be calculated using the first passage time theory(38-40). The barriers for a diffusing head to reach a binding site on MT were given by the configurational computation (see the RESULTS AND DISCUSSIONS section), and the barriers enter the calculation of the first passage times(38-40).

The kinetic Monte Carlo simulation yielded a temporal series of mechanochemical states and positions of the two heads as the dimer makes steps along MT. When a double-headed binding configuration occurs, the positions of both heads along MT are recorded. Upon hydrolysis-initiated detachment of a head from MT, the position of the mobile head is updated to be its average position during random diffusion. Because the linker chains restrict the mobile head's diffusion within a sphere centered at the other, MT-bound head, the average position of the diffusing head along MT is just the position of the standing head.

The simulation used measured values for the enzymatic rates of catalytic cores. The following rates were taken from ref. (41) and references therein: ATP binding rate 3 μM s$^{-1}$, reverse dissociation rate 150 s$^{-1}$, hydrolysis rate 200 s$^{-1}$, rate for reverse ATP synthesis 25 s$^{-1}$, rate for γ-phosphate release 250 s$^{-1}$. Diffusion coefficient for head diffusion was taken as $3.5 \times 10^6$ nm$^2$/s, which is 1-2 orders of magnitude lower than values found experimentally for intra-chain diffusion of bare，short peptides(42-44). We assumed in the simulation that the enzymatic rates of a motor domain are affected by a rear-pointing force inflicted upon it by its adjacent linker peptide. Specifically, we assumed that the ATP hydrolysis rate ($k_{hyd}$) and the ATP dissociation rate ($k_{off}$) depend on the rear-pointing force (*F*) by a Boltzmann-type relationship

$$k_{hyd}(F) = k_{hyd}(F=0)/[p_1 + q_1 \exp(F\delta/k_BT)], \quad (5)$$

$$k_{off}(F) = k_{off}(F=0)/[p_2 + q_2 \exp(F\delta/k_BT)], \quad (6)$$

where $p_1 + q_1 = p_2 + q_2 = 1$. We assumed $p_1 = p_2$, and used parameters deduced by Schnitzer, Visscher and Block in Ref.(5): $\delta = 3.7$ nm and $q_1 = 0.0062$. We note that *F* entering the above equations is the calculated force imposed upon the motor domain by its adjacent linker, and is not equal to the external load applied to the stalk domain for a double-headed kinesin-MT binding configuration.

In the kinetic Monte Carlo simulation ATP diffusively binds to a MT-bound, nucleotide-free head and the ATP binding results in linker zippering. The zippered conformation of the neck linker is maintained through the subsequent hydrolysis ATP + K + M → ADP•Pi + K + M until product release (here K denotes kinesin head and M denotes MT). Release of γ-phosphate (Pi) from the catalytic core triggers detachment of the ADP-carrying head off MT. For double-binding configurations, detachment of the ADP-associated head is likely assisted by the mechanical strain of the neck linkers(1). ADP release from a head is assumed to occur upon its binding to MT. Selection rules derived from the configurational analysis were implemented into the simulation, e.g. transitions are forbidden if the difference in configurational energies is higher than energy available from ATP hydrolysis.

**RESULTS AND DISCUSSIONS**

**Major kinesin-MT binding configurations**

The major dimer-MT configurations to be considered are schematically illustrated in insets of Fig. 1 B. In single-headed binding states the standing head either binds ATP or not (marked as state II and I respectively. See illustrations in the figure). In double-headed binding state III (VI) only the rear (front) head binds ATP with the adjacent linker being zippered. In state IV (V) both heads are nucleotide-free (ATP-bound). Transitions between the double-headed binding states (and between the single-headed ones) occur via diffusive binding of ATP to the catalytic core within a head or reverse ATP dissociation. Transitions between double-headed and single-headed bindings occur by search-and-binding of a diffusing head to MT and reverse head detachment.

**Zippering-facilitated diffusional bias**

During kinesin's steps a mobile head reaches a nearby binding site on MT via intra-chain diffusion in which the linkers are self-stretched. The self-stretching drains conformational entropy out of the linker chains, and causes a free-energy barrier for kinesin's steps. Linker zippering at the standing head points the diffusing head towards the binding site to the MT plus end, thus reducing the barrier for forward steps(11,13). The lowest barriers for forward and backward steps were calculated using eq. 1. More specifically the lowest barrier for a forward step occurs when the kinesin-MT system undergoes transition from configuration II to III (see illustrations in Fig. 1 B), and the barrier is quantitatively given by $F_F = F(III) = E(III) - E(II) - U_B$. Here $F(III)$ is the Helmholtz free energy of the linker chains in configuration III as given by eq. 1, $E(II)$ and $E(III)$ are configurational energies for configuration II and III as presented in Fig. 1 B. Similarly the lowest barrier for backward steps occurs when the kinesin-MT system undergoes transition from configuration I to IV backwardly, and the corresponding barrier is $F_B = F(IV) = E(IV) - E(I) - U_B$. In the equations for both barriers $U_B$ takes the value for a nucleotide-free head. Origin of the barriers is the intra-chain potential that acts against self-stretching of the linker peptides during the mobile head's diffusive search for a binding site.

Fig. 1A presents the results as a function of trial values for zippered length. For both barriers, satisfactory agreement with measured values(14) occurs at $n_Z$ = 7 residues for zippered length. This value lies within the range of 5 – 10 residues deduced from mutagenesis studies(11-13), and will be used throughout the present study. The measured and predicted values for barrier difference between forward and backward steps are both ~ 6 $k_B T$. Thus the small zippering energy (~ 1.2 $k_B T$) is amplified into a much larger diffusional bias. This zippering-facilitated diffusional bias is in spirit of the well-studied Brownian motor mechanism(45,46). We note however that the diffusional bias of ~ 6 $k_B T$ is readily compromised by opposing loads as small as 2 pN, which is far below the observed stalling forces.

**Removal of dimer-MT configurational degeneracy and onset of asymmetric ground state**

A major finding of this study is that overstretching of the linker peptides re-organizes kinesin-MT binding configurations into a unique energy hierarchy, which in turn facilitates a directional locking in addition to diffusional bias. This novel role of neck linkers is illuminated by considering configurational energies as a function of hypothetical length change of linker peptides (Fig. 1 B).

Let us first consider the hypothetic case in which the linker length is much larger than the binding site period of MT (i.e. $l_N >> d$) so that the free energy of both linker chains is negligible (i.e. $F \to 0$). In such a long-linker limit state V would be lowest in energy, and VI and III be degenerate (i.e. equal in energy). The ground state offers no directional preference for the dimer's movement, because both heads adopt the same mechanochemical state. States III and VI are inversely asymmetric in terms of mechanochemical states of the heads, but both states occur with equal chance according to Boltzmann's law canceling any net directional preference. Thus the overall dimer-MT interacting dynamics is directionless.

As the linker length approaches the binding site period, mechanical strain of the linkers raises energies of state III – VI to differing degrees depending on their internal geometry according to eq. 1 – 3. This mechanical effect causes re-ordering of configurational energies and removes the configurational degeneracy. Over the linker length range of $n_N$ ~ 12 – 50 residues, state V and VI are elevated in energy, but state III is less affected and becomes the new asymmetric ground state for the dimer-MT system.

Degeneracy removal and onset of a unique asymmetric ground state are the basis for kinesin's unidirectionality. The configurational hierarchy shown in Fig. 1 B exposes two distinct regimes for rectification of directional movement for a kinesin dimer.

**The regime of probabilistic bias**

The first regime corresponds to $n_N$ ~ 22 – 50 residues, in which the asymmetric ground state occurs and the double-binding states III – VI all have energies below those of single-binding states. Disruption of the double-headed binding states then requires energy input, which is supplied by ATP hydrolysis at a motor head. An ATP- and MT-bound head is on the pathway toward active detachment of the head from MT, which is triggered by post-hydrolysis phosphate release(47). Therefore the kinesin-MT states III, V and VI are all transient states to be disrupted by detachment of their ATP-bound heads. Accordingly, in the single-zippering state III the rear head is readily detached. After a diffusion process the mobile head may re-bind to MT either at the previous position or at the binding site before the standing head. The dimer thus makes a forward step or stays. A backward step is impossible directly from state III, because the front head has no energy supply for active detachment. The other single-zippering state VI is readily disrupted by hydrolysis at the front head, allowing a backward step but not a forward one.

As a high-energy state a single-headed binding can decay to any of the double-headed states III – VI. Through repeated cycles of hydrolysis-powered disruption and spontaneous regeneration of double-headed bindings, occurrences of state III tend to cause forward steps and occurrences of state VI tend to cause backward steps. The forward preference will prevail over the backward one, because state III, being the ground state for kinesin-MT system, occurs with a higher probability than state VI. Because of the small energy gap between both states, the net directional bias occurs in an average sense.

The range of $n_N$ ~ 22 – 50 residues is therefore a regime of probabilistic bias.

Processivity is poor due to occurrences of the symmetric state V, whose two ATP-bound heads can be detached simultaneously to throw the entire dimer off MT.

**The regime of decisive directional locking**

Normally functioning kinesin dimers lie in the second regime of $n_N$ ~ 12 – 21 residues. In this regime vast energy-gaps occur which are insurmountable by the energy released from ATP hydrolysis (~ 25 $k_B T$). Consequently states V and VI become forbidden states, whilst state III remains as the ground state. In absence of state VI and V, ATP-powered detachment occurs only for state III, and invariably for the rear head. After each hydrolysis event a dimer's center of mass moves forwards or stays, but never turns back. No consecutive backward steps can develop through repeated hydrolysis cycles. Linker shortening thus transforms the probabilistic bias into a direction-locking effect of deterministic nature. Exclusion of state V suppresses concurrence of hydrolysis-facilitated detachment of both heads, and drastically extends the dimer's run length.

Within the direction-locking regime, ATP binding to the rear head of state IV is favored, because this brings the dimer-MT system to the ground state. However, ATP binding to the front head of state IV, and also of the ground state III, is energetically prohibited, because ensuing linker zippering amounts to transition to inaccessible state VI or V. Thus the same head is allowed to accept ATP in a trailing position, but not in a leading position. After a head consumes ATP and successfully binds to MT in front of the other standing head, the newly settled head loses its ATP-accepting status to the head that now lies behind. Such a position-dependent head-head coordination ensures that a kinesin dimer runs in a head-over-head gait with the two heads hydrolyzing ATP alternately. The system-level transition rules thus fashion local conformational change (linker zippering) into long-range head-head coordination.

**Load-bearing capacity of kinesin dimers**

Load-bearing capacity of a kinesin dimer can be quantified by load-deformed dimer-MT configurational hierarchy. The results in Fig. 1C shows that both states III and IV remain lower in energy than single-headed binding states up to an opposing force of ~ 8 pN, which coincides with the upper limit of measured stall forces(3). Below this threshold force, head detachment is only possible with energy supplied from ATP hydrolysis and the directional locking preserves. This explains the early observation(3) that kinesin dimers develop no consecutive steps up to stall forces of ~ 8 pN. Above the threshold force state IV becomes higher in energy than single-headed binding states. Unstable state IV will decay to single-headed binding by spontaneous detachment of the load-bearing front head, rendering defect the directional locking. Occurrences of state IV by a load-directed backward binding from state I make possible consecutive back-steps. This rationalizes the recent finding of consecutive back-steps under super-stall forces of ~ 10 – 15 pN(4). When the force further increases to ~ 19 pN, even state III becomes unstable rendering the directional locking groundless completely.

**Conventional Kinesin is molecular ratchet-and-pawl device**
The direction-locking capability indicates that conventional kinesin is essentially a molecular ratchet-and-pawl device. One may regard as "ratchet" the asymmetric ground state, in which the two identical heads adopt different mechanochemical states depending on their being in the leading or trailing position with respect to the MT plus end. The "pawl" is hydrolysis-powered selective detachment of the rear head but not the leading head in a dounble-headed dimer-MT binding state. We note that the position-dependence of head states in the ground state (i.e. the ratchet) is the basis for the discriminate head detachment (i.e. the pawl). The ratchet-and-pawl device functions most ideally in the direction-locking regime mentioned before. In the bias regime the ratchet (i.e. asymmetric ground state) is preserved, but the pawl is defect because hydrolysis-enabled detachment of the rear head is not completely impossible.

Such a synergic ratchet-and-pawl mechanism is the unified physical mechanism for conventional kinesin's directionality and processivity. First, it is this rather load-insensitive ratchet-and-pawl mechanism that selects the direction coincided with orientation of zippered neck linkers and locks the dimer's movement into it in defiance of even stalling loads. The zippering-biased diffusion, being susceptible to loads, merely reinforces the directionality by promoting occurrence of successful steps. Second, the ratchet-and-pawl mechanism enables a dimer to walk consecutive steps as long as ATP turn-over rate at a MT-bound head is much lower than the rate for diffusive search-and-binding of the other head. Enzymatic rates determined experimentally and diffusion times calculated with barriers from the configurational computation show that the above time requirement is satisfied by kinesin.

The configurational hierarchy in the direction-locking regime clarifies the mechanochemical cycle for kinesin's steps. As can be seen in Fig. 1B, only three categories of dimer-MT binding states are accessible, i.e. the ground state (state III), single-binding ones (state I and II) and zippering-free double-binding state (state IV). The ground state and the two single-binding states form the major mechanochemical cycle for kinesin's steps (see illustration Fig. 2 A). State IV rarely occurs at low loads because reaching it by forward or backward binding from state I encounters a barrier of ~ 15 $k_B T$.

**A molecular-mechanical basis for the kinetic model of alternate head catalysis**
By linker zippering multiple molecular contacts are formed between a catalytic core domain and the adjacent linker peptide. Both enthalpy and entropy changes are rather large (~ 50 $k_B T$)(13), although the net change in free energy is small. Such an extensive linker-catalytic core binding likely causes structural adjustment inside the catalytic core domain in addition to the conformational change of the liker peptide. On one hand it is known that ATP binding to a catalytic core initiates the linker-catalytic core binding, which is maintained until post-hydrolysis phosphate release(11). On the other hand, nucleotide processes at the catalytic core may in turn be affected by the ensuing structural change within the catalytic core domain. A possible scenario is that the zippering-facilitated structural change within the catalytic core domain is required for stable ATP binding or /and subsequent hydrolysis reaction. Then frustrated zippering at the leading head in a double-headed dimer-MT binding state will cause the ATPase cycle at the front head to lag behind that at the rear head. The insurmountable energy gaps in the kinesin-MT configurational hierarchy ensures a sufficiently large rearward strain, which prohibits linker zippering at a leading head and thereby postpone its ATP consumption. This is in line with recent experimental studies(5,48,49) that suggested reduced nucleotide affinity of a catalytic core's active sites under rearward strain. Thus the transition rules in the direction-locking regime provide a molecular mechanical basis for the kinetic model of alternate head catalysis(7-9).

**Dynamical simulation supports the ratchet-and-pawl mechanism**
The kinetic Monte Carlo simulation of dynamical transitions between dimer-MT configurations produces trajectories of processive walking in a hand-over-hand gait. Fig. 2 B presents time evolution of positions of both heads along MT from a typical run of the dimer as found in the simulation. The opposing load is as high as 5.6 pN. In interpreting the results it is important to note that Fig. 2 B shows for a diffusing head its average position, which is identical to the position of the other MT-bound head. Thus in Fig. 2 B hydrolysis-initiated detachment of a rear head is shown as a ~ 8 nm advance of the head's position. Similarly, a diffusing head's successful binding to MT at a forward site or a backward site is represented by a ~ 8 nm advance or retreat of the head's position. Therefore, the ~ 8 nm change of a head's position as shown in Fig. 2 B indicates normal detachment or attachment events rather than any sub-steps. The entire dimer makes a full step of ~ 8 nm when a rear head is detached upon post-hydrolysis phosphate release and then binds over a distance of ~ 16 nm to a forward site. The trajectories in Fig. 2 B show however that the mobile head can keep diffusing for a long time because of reduced search-and-binding rates under the close-to-stall load. Consequently a full step of ~ 16 nm is often found in the figure as two ~8 nm advances separated by a long-lived diffusing state. A close look of both heads' trajectories reveals that the directional bias caused by linker zippering at the standing head is defiled by the close-to-stall load because the diffusive head frequently binds back to its former binding site on MT. But the locking mechanism prevents these individual back-bindings from developing into consecutive back-steps, instead the dimer's center-of-mass maintains an intermittent procession of forward steps.

Using center-of-mass trajectories of kiensin dimers $x(t)$ generated by the simulation,

we have calculated conventional kinesin's average velocity and stepping irregularity. The stepping irregularity is quantified by the randomness parameter, *r*, which is defined as (50)

$$r = \lim_{t \to \infty} \frac{\langle x^2(t) \rangle - \langle x(t) \rangle^2}{d \langle x(t) \rangle}$$ (angle brackets denote ensemble average). Values of the

randomness parameter also serve as an indicator for ATP consumption during kinesin's steps. As can be seen in Fig. 2 C-F the simulation results satisfactorily reproduce measured velocity and stepping irregularity for a broad range of opposing loads and ATP concentrations(3,50). The stall forces from the simulations are also close to measured values of 5-8 pN(3) depending on ATP concentrations (Fig. 2 D). These results confirm that the ratchet-and-pawl plus bias mechanism works in kinesin.

The present theory predicts a tight coupling between kinesin's ATPase pathway and mechanical movement in agreement with experimental findings(50,51). The coupling ratio, i.e. the average number of ATP molecules hydrolyzed per forward step was given by the simulation straightforwardly, because our simulation kept a record of ATP consumption and kinesin steps for each run. The simulation yielded a value of 1 for the coupling ratio up to load of 5 pN (Fig. 2 D). Consistent with this result, the randomness parameter remains to be 0.5 up to 5 pN in both the measured data and the simulation results. As the load is further increased, the measured randomness data are underestimated by the simulation results, probably because the present simulations neglect load sensitivity of some transitions in the mechanochemical cycle. At such extreme loads, the randomness analysis might become invalid due to loss of processivity(3,50).

**Kinesin's ratchet-and-pawl device is evolutionarily optimized**

The synergetic ratchet-and-pawl mechanism arises from a fine interplay of multiple molecular properties of kinesin-MT system, which include not only the linker length but also the catalytic core's capabilities for MT binding and for nucleotide-dependent linker zippering. Configurational computations in which these properties are hypothetically changed provide a quantitative basis for assessing how well kinesin is evolutionarily adapted to its motor function. The ideal working regime for the ratchet-and-pawl mechanism, namely the direction-locking regime mentioned before, can be quantitatively defined by two key requirements: thermodynamic stability of the asymmetric ground state (state III) and inaccessibility of state V and VI enforcing the forbidden transitions. Both requirements yield respectively lower and upper boundaries for the regime in terms of effective linker length in double-headed dimer-MT bindings. As shown in Fig. 1 B, the lower boundary $n_1$ is approximately given by energy-level crossing between the lowest-lying single-binding state (state II) and the ground state (state III), and the level crossing between state II and the double-binding double zippering state (state V) gives the upper boundary $n_2$. Fig. 3 presents values of $n_1$ and $\Delta n = n_2 - n_1$ for hypothetic variance of zippered length ($n_Z$) between 1-10 amino acid residues, free-energy gain by linker zippering ($U_Z$) between 0.5-7 $k_B T$, and head-MT binding energy for a ATP-bound or nucleotide-free head ($U_{KM}$) between 10-20 $k_BT$ (The binding energy for an ATP-bound head was found to be close to that for a nucleotide-free head(22,52). We assumed both binding energies to be equal in this study). Remarkably, even for such unrealistically broad change of molecular properties kinesin's effective linker length of 14-16 amino acid residues (for double-headed bindings with MT) invariably lies within the ideal regime. Kinesin appears to lie closer to the lower than the upper boundary of the working regime. This feature is likely advantageous for kinesin's motor function, because it allows the ratchet-and-pawl mechanism to function properly even when the coiled coils unwind to a non trivial extent. Overall kinesin tends to optimize robustness of its ratchet-and-pawl device against variance in effective linker length by minimizing lower boundary of the regime ($n_1$) and simultaneously maximizing its size ($\Delta n$). This joint optimization requires larger values for $U_{KM}$ and $n_Z$, both of which however have their own limits. The zippered length is restricted by the size of the catalytic core, while MT-bindings must not compromise sufficiency of ATP hydrolysis for their disruption. With an experimentally measured value for $U_{KM}$ between 16 - 19 $k_BT$ (22,52) and a likely value for $n_Z$ close to 7 amino acid residues (see Fig. 1 A), kinesin has largely approached the natural limits for both quantities. The results in Fig. 3, and also those in Fig. 1B clearly demonstrated that kinesin's necklinker length, zippering and MT-binding capabilities have been evolutionarily fine-tuned to maximize robustness of the inherent ratchet-and-pawl device, thereby ensuring the motor's sturdy directionality and high processivity. Interestingly, both boundaries of the ratchet-and-pawl regime, while being rather sensitive to the zippered length ($n_Z$), are shifted less than 1 amino acid residue by a tenfold change of the zippering energy ($U_Z$). This notable insensitivity ensures adequacy of the surprisingly small zippering energy of $U_Z \approx 1.2$ $k_BT$ found for kinesin(13), as far as the ratchet-and-pawl mechanism is concerned.

**A framework for quantitative analysis of kinesin mutants**
The ideal regime together with the two regions sandwiching it (see Fig. 1 B) provides a basis for analyzing performance of genetically engineered constructs of dimeric kinesin(11-13,53). In the regime below $n_1$ the lowest-energy state for dimer-MT double bindings even surpasses energies of single-binding states, and if processive walking is still possible depends on lifetime of the double-binding state and on the barrier for reaching it. On the other side beyond $n_2$ is the bias regime in which direction-locking is defect and processvity reduced, but net plus-end directionality survives albeit in an average sense. In which regime a mutant dimer actually lies depends not only on effective linker length, but also on other molecular properties which participate in defining boundaries of the regimes. An example is a study in which 6 or 12 amino acids were inserted into the junction of the neck linker and coiled coil(53). Constructs from this study, though having considerably elongated neck linkers, fall into the regime of defect but not abolished ratchet-and-pawl, which explains the observation of shortened run length, reduced velocity and survival of averaged direction towards MT plus end(53).

As the unified molecular-physical mechanism for conventional kinesin's directionality and processivity, the synergic ratchet-and-pawl mechanism establishes a quantitative link between the motor's overall performance and a list of well-defined molecular properties of the dimer-MT system. This provides a tool for rational design for mutations for future studies on kinesin, and also for study of hereditary mutations involved in human neurodegenerative diseases(54-56).

**CONCLUSIONS**
In conclusion, conventional kinesin is an evolutionarily fine-tuned molecular ratchet-and-pawl device that locks its movement into a unique polarity of MT and ensures consecutive steps in a head-over-head gait. Remarkably, this conclusion was established by the system-level configurational analysis employing merely first principles plus several measured binding and barrier energies. The conclusion is supported by a simulation study for the dimer's running process based on the identified ratchet-and-pawl mechanism and incorporating measured enzymatic rates. The present findings quantitatively rationalize a large body of previously puzzling results. The load-insensitive direction-locking by the ratchet-and-pawl mechanism explains kinesin's unyielding direction, which the zippering-induced diffusional bias merely reinforces. When the load reaches a certain super-stall range, kinesin's ratchet-and-pawl device becomes defect but not entirely abolished. As a consequence, rare events of consecutive back-stepping occur as seen experimentally. Kinesin's ratchet-and-pawl is notably insensitive to the zippering energy, rationalizing its small value found experimentally. The method of system-level energetic computation and analysis introduced in this work has turned out to be powerful in exposing synergetic molecular mechanisms. The method may also be useful in study of other processive motor proteins such as myosin V and cytoplasmic dynein.

**APEENDIX: ATOMIC COMPUTATION FOR LINKER PEPTIDES**
Bending rigidity and loop-forming property of a neck linker, both being important in determining the free energy of the linker chains, depend on atomic details of the linker peptide, particularly its backbones. We used poly-alanines as a model for linker peptide of hypothetically varying length. To obtain reliable values for $l_p$ and $l_T$, we used the fast pivot Monte Carlo procedure based on an all-atom representation(44) to generate an extremely large ensemble of peptide conformations. The conformational ensemble yield average radius of gyration as a function of peptide length in good agreement with experimental data(57). The average end-to-end distance calculated from the sampled conformations yields persistence length through the formula of worm-like chain model, $\langle R^2 \rangle = 2l_p l_N - 2l_p^2 [1 - \exp(-l_N / l_p)]$. The procedure leads to a stable value of $l_p$ = 0.8 nm over the length range of 10-20 amino acid residues relevant to kinesin's linkers. The deduced $l_p$ value is close to those found by single-molecule measurements(30,31). A closed loop is defined by end-to-end distance smaller than 0.4 nm, and our computer-generated ensemble of peptides turned out to be sufficient for reliable computation for the probability of loop formation over seven orders of magnitude. The loop-formation probability as a function of number of amino acid residues exhibits a peak at 4 amino acids, and drops by several orders of magnitude at 2 amino acids in consistency with prediction of polymer theory(26). The minimum length for a half loop in kinesin-MT configurations is taken as 1 amino acid, yielding $l_T$ = 0.36 nm.

This work was partly funded by National Natural Science Foundation of China (Grant No. 90403006), Chinese Ministry of Education (under Program for New Century Excellent Talents in University), Shanghai Education Development Foundation (under Shuguang Program), and Shanghai Pujiang Program.

**REFERENCES**

1. Vale, R. D., and R. A. Milligan. 2000. The way things move: Looking under the hood of molecular motor proteins. Science 288:88-95.
2. Coppin, C. M., D. W. Pierce, L. Hsu, and R. D. Vale. 1997. The load dependence of kinesin's mechanical cycle. Proc. Natl. Acad. Sci. USA 94:8539-8544.
3. Visscher, K., M. J. Schnitzer, and S. M. Block. 1999. Single kinesin molecules studied with a molecular force clamp. Nature 400:184-189.
4. Carter, N. J., and R. A. Cross. 2005. Mechanics of the kinesin step. Nature 435:308-312.
5. Schnitzer, M. J., K. Visscher, and S. M. Block. 2000. Force production by single kinesin motors. Nat. Cell. Biol. 2:718-723.
6. Okada, Y., H. Higuchi, and N. Hirokawa. 2003. Processivity of the single-headed kinesin kif1a through biased binding to tubulin. Nature 424:574-577.
7. Hackney, D. D. 1994. Evidence for alternating head catalysis by kinesin during microtubule-stimulated atp hydrolysis. Proc. Natl. Acad. Sci. USA 91:6865-6869.
8. Ma, Y. Z., and E. W. Taylor. 1997. Interacting head mechanism of microtubule-kinesin atpase. J. Biol. Chem. 272:724-730.
9. Hancock, W. O., and J. Howard. 1999. Kinesin's processivity results from mechanical and chemical coordination between the atp hydrolysis cycles of the two motor domains. Proc. Natl. Acad. Sci. USA 96:13147-13152.
10. Yildiz, A., M. Tomishige, R. D. Vale, and P. R. Selvin. 2004. Kinesin walks hand-over-hand. Science 303:676-678.
11. Rice, S., A. W. Lin, D. Safer, C. L. Hart, N. Naber, B. O. Carragher, S. M. Cain, E. Pechatnikova, E. M. Wilson-Kubalek, M. Whittaker, E. Pate, R. Cooke, E. W. Taylor, R. A.

Milligan, and R. D. Vale. 1999. A structural change in the kinesin motor protein that drives motility. Nature 402:778-784.
12. Case, R. B., S. Rice, C. L. Hart, B. Ly, and R. D. Vale. 2000. Role of the kinesin neck linker and catalytic core in microtubule-based motility. Current Biology 10:157-160.
13. Rice, S., C. Y., C. Sindelar, N. Naber, M. Matuska, R. D. Vale, and R. Cooke. 2003. Thermodynamics properties of the kinesin neck-region docking to the catalytic core. Biophys. J. 84:1844-1854.
14. Taniguchi, Y., M. Nishiyama, Y. Ishii, and T. Yanagida. 2005. Entropy rectifies the brownian steps of kinesin. Nature Chem. Biol. 1:342-347.
15. Derenyi, I., and T. Vicsek. 1996. The kinesin walk: A dynamic model with elastically coupled heads. Proc. Natl. Acad. Sci. USA 93:6775-6779.
16. Fisher, M. E., and A. B. Kolomeisky. 1999. The force exerted by a molecular motor. Proceedings of the National Academy of Sciences of the United States of America 96:6597-6602.
17. Stratopoulos, G. N., T. E. Dialynas, and G. P. Tsironis. 1999. Directional newtonian motion and reversals of molecular motors. Phys. Lett. A 252:151-156.
18. Astumian, R. D., and I. Derenyi. 1999. A chemically reversible brownian motor: Application to kinesin and ncd. Biophys. J. 77:993-1002.
19. Fisher, M. E., and A. B. Kolomeisky. 2001. Simple mechanochemistry describes the dynamics of kinesin molecules. Proc. Natl. Acad. Sci. USA 98:7748-7753.
20. Fisher, M. E., and Y. C. Kim. 2005. Kinesin crouches to sprint but resists pushing. Proceedings of the National Academy of Sciences of the United States of America 102:16209-16214.
21. Shao, Q., and Y. Q. Gao. 2006. On the hand-over-hand mechanism of kinesin. Proc. Natl. Acad. Sci. USA 103:8072-8077.
22. Okada, Y., and N. Hirokawa. 2000. Mechanism of the single-headed processivity: Diffusional anchoring between the k-loop of kinesin and the c terminus of tubulin. Proc. Natl. Acad. Sci. USA 97:640-645.
23. Ray, S., E. Meyhofer, R. A. Milligan, and J. Howard. 1993. Kinesin follows the microtubules protofilament axis. J. Cell Biol. 121:1083-1089.
24. Hoenger, A., M. Thormahlen, R. Diaz-Avalos, M. Doerhoefer, K. N. Goldie, J. Mueller, and E. Mandelkow. 2000. A new look at the microtubule binding patterns of dimeric kinesins. J. Mol. Biol. 297:1087-1103.
25. Wade, R. H., and F. Kozielski. 2000. Structural links to kinesin directionality and movement. Nature Struct. Bio. 7:456-460.
26. Shimada, J., and H. Yamakawa. 1984. Ring-closure probabilities for twisted wormlike chains. Application to DNA. Macromolecules 17:689-698.
27. Bustamante, C., J. F. Marko, E. D. Siggia, and S. Smith. 1994. Entropic elasticity of l-phage DNA. Science 265:1599-1600.
28. Marko, J. F., and E. D. Siggia. 1995. Stretching DNA. Macromolecules 28:8759-8770.
29. Rubinstein, M., and R. H. Colby. 2003. Polymer physics. Oxford: Oxford University Press.
30. Kellermayer, M. S. Z., S. B. Smith, H. L. Granzier, and C. Bustamante. 1997. Folding-unfolding transitions in single titin molecules characterized with laser tweezers.

Science 276:1112-1116.
31. Rief, M., M. Gautel, F. Oesterhelt, J. M. Fernandez, and H. E. Gaub. 1997. Reversible unfolding of individual titin immunoglobulin domains by afm. Science 276:1109-1112.
32. Bouchiat, C., M. D. Wang, J.-F. Allemand, T. Strick, S. M. Block, and V. Croquette. 1999. Estimating the persistence length of a worm-like chain molecule from force-extension measurements. Biophys. J. 76:409-413.
33. Makarov, D. E., Z. Wang, J. B. Thompson, and H. Hansma. 2002. On the interpretation of force extension curves of single protein molecules. J. Chem. Phys 116:7760-7764.
34. Tomishige, M., and R. D. Vale. 2000. Controlling kinesin by reversible disulfide cross-linking: Identifying the motility-producing conformational change. Journal of Cell Biology 151:1081-1092.
35. Wang, Z. S., and D. E. Makarov. 2003. Nanosecond dynamics of single polypeptide molecules revealed by photoemission statistics of fluorescence resonance energy transfer: A theoretical study. J. Phys. Chem. B 107:5617-5622.
36. Metiu, H., Y. T. Lu, and Z. Zhang. 1992. Epitaxial growth and the art of computer simulations. Science 255:1088.
37. Makarov, D. E., H. Hansma, and H. Metiu. 2001. Kinetic monte carlo simulation of titin unfolding. J Chem Phys 114:1-11.
38. Szabo, A., K. Schulten, and Z. Schulten. 1980. First passage time approach to diffusion controlled reactions. J. Chem. Phys 72:4350-4357.
39. Wang, Z. 2004. A bio-inspired, laser operated molecular locomotive. Phys. Rev. E 70:031903.
40. Howard, J. 2001. Mechanics of motor proteins and the cytoskeleton. Sunderland, Massachusetts: Sinauer Associates, Inc.
41. Cross, R. A. 2004. The kinetic mechanism of kinesin. Trends Biochem. Sci. 29:301-309.
42. Buckler, D. R., E. Haas, and H. A. Scheraga. 1995. Analysis of the structure of ribonuclease a in native and partially denatured states by time-resolved nonradiative dynamic excitation energy transfer between site-specific extrinsic probes. Biochemistry 34:15965-15978.
43. Lapidus, L. J., W. A. Eaton, and J. Hofrichter. 2001. Dynamics of intramolecular contact formation in polypeptides: Distance dependence of quenching rates in a room-temperature glass. Physical Review Letters 87:258101.
44. Wang, Z. S., and D. E. Makarov. 2002. Rate of intramolecular contact formation in peptides: The loop length dependence. J. Chem. Phys 117:4591-4593.
45. Astumian, R. D. 1997. Thermodynamics and kinetics of a brownian motor. Science 276:917-922.
46. Julicher, F., A. Ajdari, and J. Prost. 1997. Modelling molecular motors. Rev. Mod. Phys. 69:1269-1281.
47. Nitta, R., M. Kikkawa, Y. Okada, and N. Hirokawa. 2004. Kif1a alternately uses two loops to bind microtubules. Science 305:678-683.
48. Rosenfeld, S. S., P. M. Fordyce, G. M. Jefferson, P. H. King, and S. M. Block. 2003. Stepping and stretching. J. Biol. Chem. 278:18550-18556.

49. Guydosh, N. R., and S. M. Block. 2006. Backsteps induced by nucleotide analogs suggest the front head of kinesin is gated by strain. Proc. Natl. Acad. Sci. USA 103:8054-8059.

50. Schnitzer, M. J., and S. M. Block. 1997. Kinesin hydrolyses one atp per 8-nm step. Nature 388:386-390.

51. Hua, W., E. C. Young, M. L. Fleming, and J. Gelles. 1997. Coupling of kinesin steps to atp hydrolysis. Nature 388:390-393.

52. Crevel, I. M. T. C., A. Lockhart, and R. A. Cross. 1996. Weak and strong states of kinesin and ncd. J. Mol. Biol. 257:66-76.

53. Hackney, D. D., M. F. Stock, J. Moore, and R. A. Patterson. 2003. Modulation of kinesin half-site adp release and kinetic processivity by a spacer between the head groups. Biochemistry 42:12011-12018.

54. Reid, E., M. Kloos, A. Ashley-Koch, L. Hughes, S. Bevan, I. K. Svenson, F. L. Graham, P. C. Gaskell, A. Dearlove, M. A. Pericak-Vance, D. C. Rubinsztein, and D. A. Marchuk. 2002. A kinesin heavy chain (kif5a) mutation in hereditary spastic paraplegia (spg10). Amer. J. Human Genet. 71:1189-1194.

55. Fichera, M., M. Lo Giudice, M. Falco, M. Sturnio, S. Amata, O. Calabrese, S. Bigoni, E. Calzolari, and M. Neri. 2004. Evidence of kinesin heavy chain (kif5a) involvement in pure hereditary spastic paraplegia. Neurology 63:1108-1110.

56. Blair, M. A., S. C. Ma, and P. Hedera. 2006. Mutation in kif5a can also cause adult-onset hereditary spastic paraplegia. Neurogenetics 7:47-50.

57. Millet, I. S., S. Doniach, and K. W. Plaxco. 2002. Towards a taxonomy of the denatured state: Small angle scattering studies of unfolded proteins. Adv. Protein Chem. 62:241-262.

**Figure Captions**

**Figure 1. Stepping barriers and configurational energies of conventional kinesin.** Kinesin-MT binding configurations are illustrated in the insets. The motor heads are represented by large symbols filled in yellow color. The ATP-bound state of a head is indicated by label T, and the ADP-bound state by label D. Unlabeled heads are nucleotide-free. The neck linkers are shown by lines in blue color, their zippered portions are shown by bold lines in red. The coiled coil dimerization domains are shown by spiral lines in cyan. The large symbols in heavy and light grey represent $\alpha$ and $\beta$ tubulin units of MT. **A**. Lowest free-energy barriers for forward and backward stepping. The filled symbols are calculated results for integer numbers of zippered amino acid residues, while the lines were drawn to guide eyes. The bias, i.e. barrier difference between forward and backward steps is also shown. The measured values for the barriers are from ref.(14). **B**. Computed energies for major kinesin-MT binding configurations as a function of hypothetically changing length of the linker peptide. The filled symbols are results for integer numbers of amino acid residues in a linker peptide, and the lines were drawn to guide eyes. Conventional kinesin's effective linker length for double-headed bindings to MT is indicated by the shadow area. **C**. Distortion of configurational hierarchy by opposing load. The shadow area indicates the measured stall forces from ref. (3), which are 5-8 pN depending on values of ATP concentrations.

**Figure 2. Walking behavior of kinesin dimers. A.** Illustration of kinesin's major mechanochemical cycle at low loads deduced from the configurational analysis (see text). The kinesin-MT system and the states of the motor heads are illustrated in the same way as in the insets of Fig. 1 **B**. At low loads three dimer-MT binding states (I-III) are likely involved. The transition from state I to II is caused by ATP binding and linker zippering at a MT-bound head. MT binding and ADP release of the diffusing head causes transition from state II to III. Hydrolysis-initiated detachment of the rear head causes transition back to state I. **B – F.** Prediction of the kinetic Monte Carlo simulation (solid lines) versus experimental data (filled symbols). **B**. Typical trajectories of both heads. Initially at zero time the two heads are both bound to MT. After hydrolysis-initiated detachment the diffusing head is allowed to bind MT again only at binding sites other than the one occupied by the standing head. Thus the head whose trajectory is shown by solid lines in red (black) binds MT only at positions indicated by red (black) dashed lines. A color mismatch between solid lines (head trajectories) and dashed lines (MT sites) indicates the diffusing state of a head. **C, D**. Average velocity of the dimer as a function of ATP concentrations and opposing loads. The measured data are from ref. (3). **E, F**. Temporal fluctuation of the dimer's walking steps as a function of ATP concentrations and loads. The measured data are from ref. (50) for **E** and ref. (3) for **F**. The overall mechanochemical coupling ratio, namely average number of ATP molecules consumed per forward step is also shown in **F**.

**Figure 3. Robustness of kinesin's ratchet-and-pawl.** Lower boundary ($n_1$) and size ($\Delta n = n_2 - n_1$) of the ideal working regime for the ratchet-and-pawl mechanism in terms of effective linker length as a function of hypothetical changes in zippered length of neck linkers upon zippering ($n_Z$), associated free-energy gain ($U_Z$), and head-MT binding energies for a nucleotide-free or ATP-bound head ($U_{KM}$) (Both binding energies were assumed equal in obtaining the results shown by the figures). For other kinesin-MT parameters the same values as for Fig. 1 are used. Definitions of $n_1$ and $n_2$ are shown in Fig. 1 **B**. The $n_1$ and $\Delta n$ values for conventional kinesin are indicated by the white areas.

Figure 1. Wang et al.

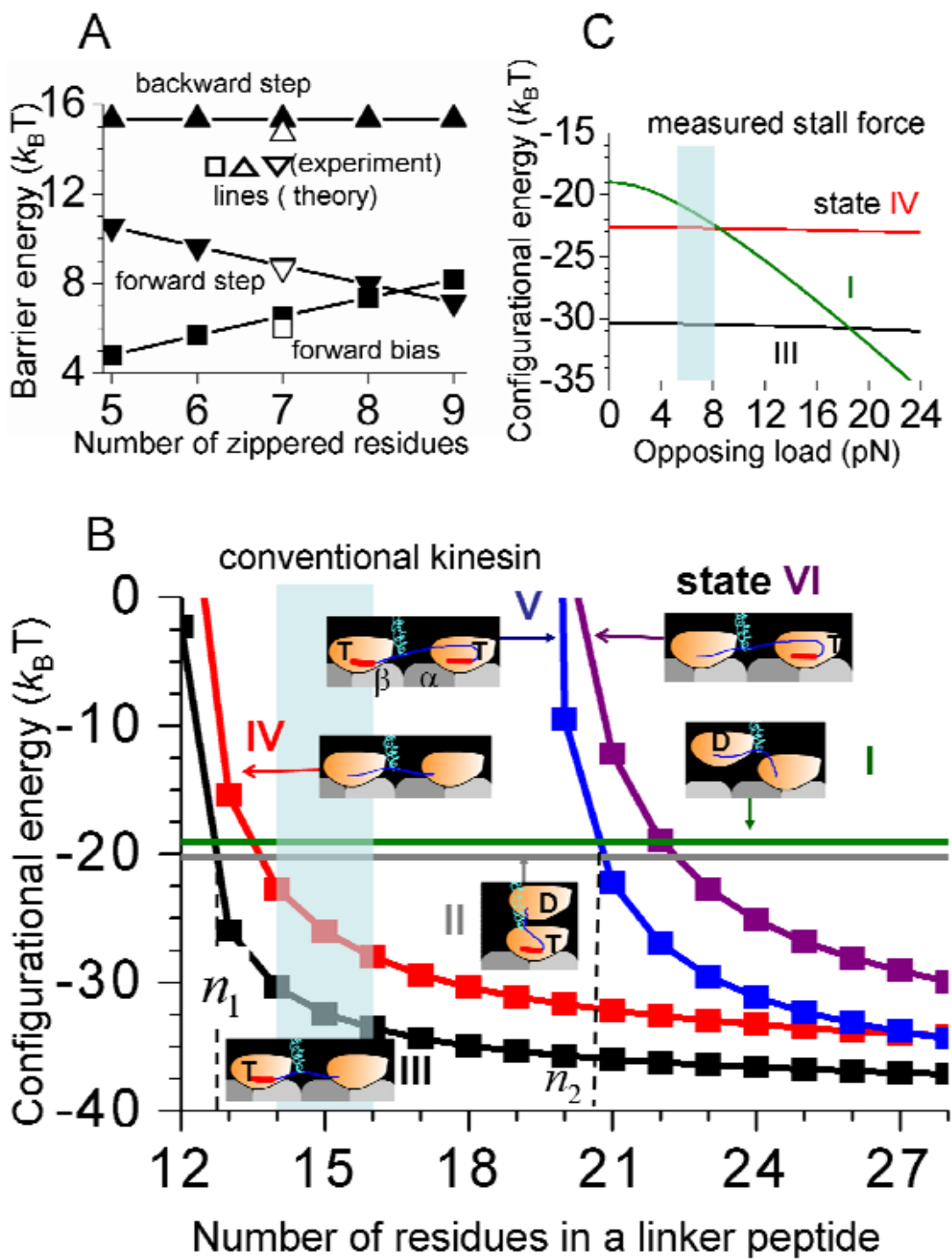

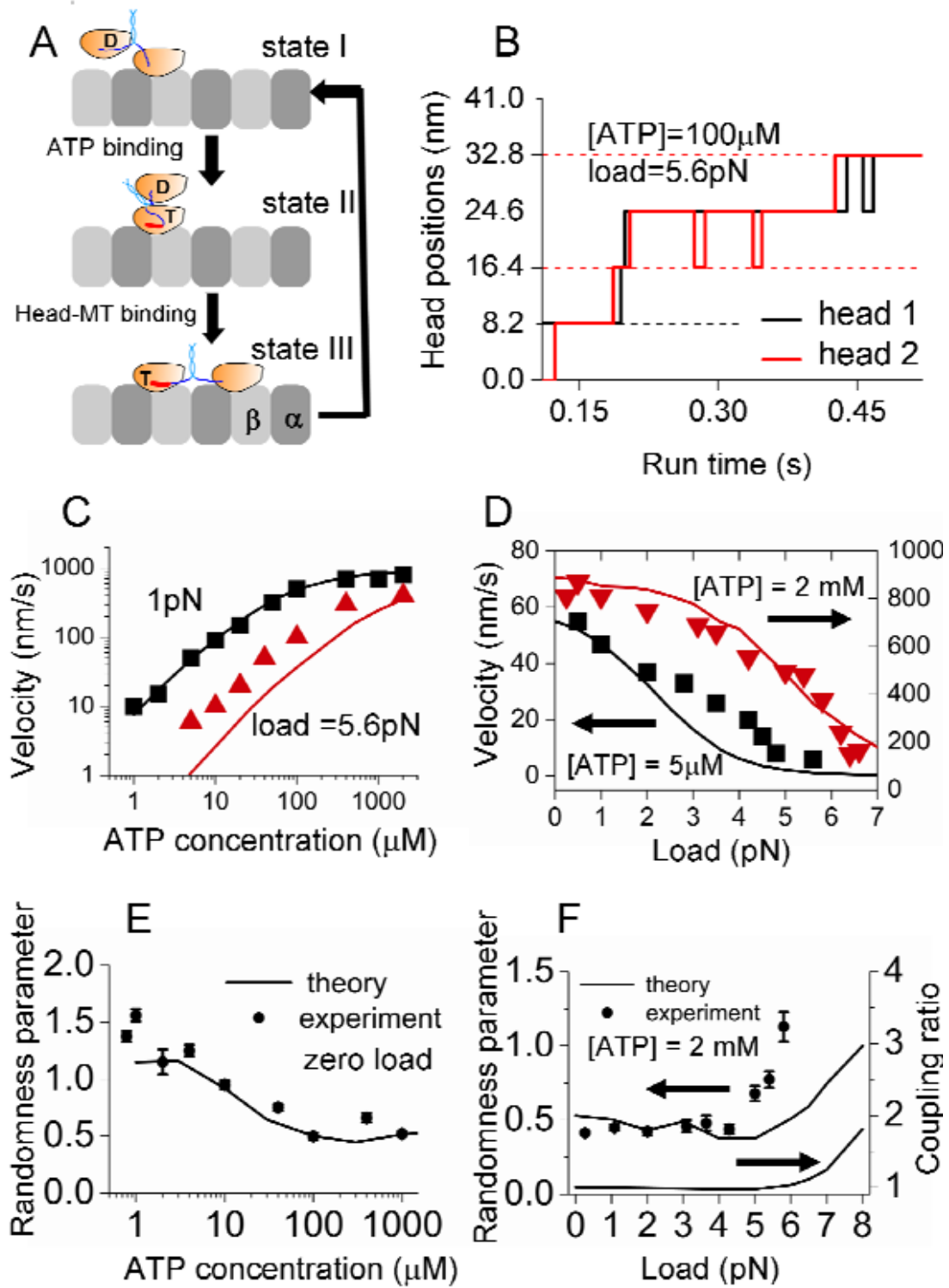

Figure 2. Wang et al.

Figure 3. Wang et al.

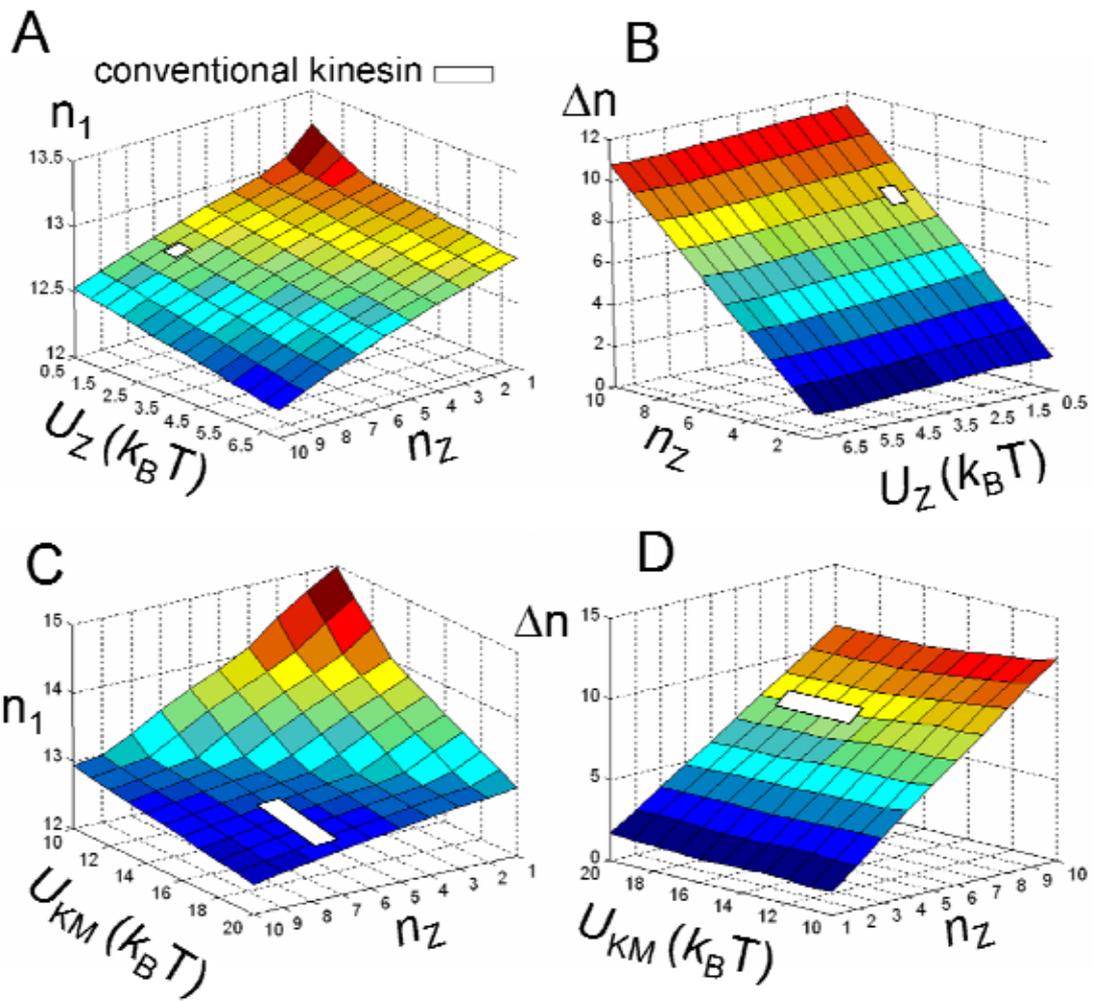